\documentclass[11pt, twopage, reprint, superscriptaddress, preprintnumbers,nofootinbib]{article}

\usepackage{amsmath}
\usepackage{amsmath}
\usepackage{amssymb}
\usepackage{mathtools,slashed}
\usepackage{physics}
\usepackage{mathtools}
\usepackage{cite}
\usepackage{bbm}
\usepackage{upgreek}

\usepackage{blindtext} 
\usepackage{bookman}
\usepackage[T1]{fontenc} 
\usepackage{microtype} 

\usepackage[english]{babel} 

\usepackage[hmarginratio=1:1,top=32mm,columnsep=20pt]{geometry} 
\usepackage[hang, small,labelfont=bf,up,textfont=it,up]{caption} 
\usepackage{booktabs} 

\usepackage{lettrine} 

\usepackage{enumitem} 
\setlist[itemize]{noitemsep} 

\usepackage{abstract} 

\usepackage{titlesec} 
\renewcommand\thesection{\Roman{section}} 
\renewcommand\thesubsection{\roman{subsection}} 
\titleformat{\section}[block]{\large\scshape\centering}{\thesection.}{1em}{} 
\titleformat{\subsection}[block]{\large}{\thesubsection.}{1em}{} 

\usepackage{titling} 

\usepackage{hyperref} 
\hypersetup{
    colorlinks=true,
    linkcolor=black,
    filecolor=magenta,      
    urlcolor=blue,
    citecolor=blue,
    pdftitle={Overleaf Example},
    pdfpagemode=FullScreen,
    }

\usepackage{physics}

\title{On the $\uptheta$-vacua and CP violation}
\author{%
\textsc{Archil Kobakhidze} 
\vspace{0.2cm} \\
\normalsize \itshape
Sydney Consortium for Particle Physics and Cosmology, \\
\normalsize  \itshape
School of Physics, The University of Sydney, NSW 2006, Australia \\ 
}
\date{} 
\renewcommand{\maketitlehookd}{%

\begin{abstract}
\noindent 
Recent claims have suggested the absence of CP violation in theories with a $\theta$-vacuum structure, particularly in quantum chromodynamics. We highlight several key points, from a perspective that is not widely discussed in the literature, which clarify why such conclusions are incorrect. In particular, an open boundary in a finite-volume theory must be accompanied by boundary degrees of freedom, the edge modes, in order to preserve large gauge invariance and faithfully capture the topological features of the theory. In the infinite-volume limit, these edge states become non-dynamical, leaving the standard $\uptheta$-vacuum structure intact, irrespective of whether this limit is taken before or after summing over topological sectors. Consequently, the $\uptheta$-vacuum structure does give rise to observable CP violation once the theory is consistently quantised.

\end{abstract}
}

\begin{document}
\maketitle


\section{Introduction}

In Refs.~\cite{Ai:2020ptm, Ai:2024cnp}, the authors argue that in non-Abelian gauge theories with a $\uptheta$-vacuum structure, most notably in quantum chromodynamics, physical observables do not depend on $\uptheta$, implying that the combined CP symmetry is conserved. If correct, this would constitute a groundbreaking result: it would resolve the long-standing strong CP problem and obviate the need for axions or other exotic physics beyond the Standard Model.

However, the results of \cite{Ai:2020ptm, Ai:2024cnp} are in contradiction with previous studies. Within the 'standard approach', the properly defined vacuum state of a non-Abelian gauge theory — that is, a gauge invariant eigenstate of the Hamiltonian $H$, is given by a superposition of topologically distinct classical vacuum states, $\lvert n\rangle$ ($n\in\mathbb{Z}$), parameterised by an angular variable $\uptheta$ ($\uptheta \sim \uptheta + 2\pi$):
\begin{equation}
\lvert \uptheta \rangle = \sum_n e^{i n \uptheta} \lvert n \rangle~.
\label{theta}
\end{equation}
The corresponding (Euclidean) generating functional of the theory (and thus any physically relevant observable),
\begin{eqnarray}
Z[\uptheta]
&\equiv&
\lim_{V,T\to\infty}
\langle \uptheta \lvert e^{-H_V T} \rvert \uptheta \rangle
\nonumber \\
&=&
\lim_{V,T\to\infty}
\sum_{n,\nu} e^{i \uptheta \nu}
\langle n-\nu \lvert e^{-H_V T} \rvert n \rangle~,
\label{ampl}
\end{eqnarray}
display explicit $\uptheta$ dependence. In particular, $\uptheta$ appears in CP-violating observables in the QCD sector of the Standard Model, giving rise to the strong CP problem.

The key technical claim emphasised in Refs.~\cite{Ai:2020ptm, Ai:2024cnp} is that the operations of summing over topological sectors $\nu \in \mathbb{Z}$ and taking the infinite four-volume limit do not commute. The authors argue that, since the integer-valued topological charge $\nu$ is only well defined in the infinite-volume limit, one must first take $V,T\to\infty$ and only then perform the sum over $\nu$. Under this prescription, the $\uptheta$-dependent phase in Eq.~(\ref{ampl}) factors out, and since $\uptheta$ then appears only as an overall phase in quantum amplitudes, it becomes unobservable. Equivalently, if one takes the limit $V,T \to \infty$ prior to summing over topological sectors, the amplitude
\begin{equation}
A(V,T, n,\nu)=\langle n-\nu \lvert e^{-H_V T} \rvert n \rangle
\label{amp}
\end{equation}
becomes independent of $\nu$ in the infinite volume limit, i.e., 
\begin{equation}
 A(n)=\lim_{V,T \to \infty}A(V,T,n,\nu)\,. 
 \label{amp2}
\end{equation}
This effectively eliminates interference between different topological sectors, rendering the CP-violating $\uptheta$ parameter redundant, $Z[\theta]\propto \sum_{\nu}{\rm e}^{i\nu\uptheta}=2\pi\delta(\uptheta)$. Given the central role of such interference in constructing a gauge-invariant vacuum and in computing physical correlators, this conclusion is highly nontrivial and requires careful scrutiny.

In what follows, we argue that a careful treatment of the theory at finite spacetime volume with open boundaries, as opposed to fixed pure-gauge boundary conditions, reveals the presence of dynamical boundary degrees of freedom.\footnote{The necessity to carefully account for additional degrees of freedom in finite-volume theories has also been argued by Dvali \cite{Dvali:2022fdv}, based on an effective three-form gauge field description of $\uptheta$ vacua.} These edge modes carry topological information and persist in the infinite-volume limit, thereby preserving interference between topological sectors and ensuring that the $\uptheta$ parameter appears in CP-violating observables.

\section{Edge modes in finite-volume gauge theory}

Let us consider a gauge theory based on the compact and simple gauge group $G$. The theory is defined on 4D Euclidean space $\mathbb{R}^4$. We partition $\mathbb{R}^4$ into two parts $M$ and $R^4/M$, with a boundary between them $\partial M$. The $G$-valued gauge potential confined to a subspace $M$ we denote as $A_{\mu}$, while $\mathcal{A}_{\mu}$ is the gauge potential defined on $R^4/M$. These two are glued on the boundary up to a gauge transform $g\in G$, i.e.,
\begin{equation}
\left.A_{\mu}\right\vert_{\partial M}=\left.g\mathcal{A}_{\mu}g^{-1}+g\partial_{\mu}g^{-1}\right\vert_{\partial M}\,.
    \label{match}
\end{equation}
The gauge potential is assumed to be a smooth function, while the gauge function $g$ is required to be smooth only on the boundary, i.e., it can be singular in the bulk. The associated degrees of freedom localised on the boundary $\partial M$ are in fact dynamical boundary degrees of freedom, known as \emph{edge modes}.

Thus, an observer in $M$ describes the above gauge theory at classical level by the following action 
\begin{eqnarray}
S&=&
\int_{M} d^4x
\left[
\frac{1}{2g^2}
\mathrm{Tr}\left(F_{\mu\nu} F^{\mu\nu}\right)
\right] \label{action}\\
&-&\int_{\partial M} \frac{d^3x}{g^2}\,
n_{\mu}
\mathrm{Tr}\left(
g^{\mu\nu}
\bigl(A_{\nu}-gD_{\nu}[\mathcal{A}]g^{-1}\bigr)
\right)\,,\nonumber 
\end{eqnarray}
where the first term is the standard Yang–Mills action. The field strength and gauge potential are defined on the four-dimensional manifold $M$, parametrised by coordinates $x_{\mu}=(\tau,x_i)$ with $i=1,2,3$,
\begin{eqnarray}
F_{\mu\nu}(\tau,x_i)
&\equiv&
F_{\mu\nu}^a T^a
\label{field}
\nonumber\\
&=&
\partial_{\mu}A_{\nu}-\partial_{\nu}A_{\mu}
-[A_{\mu},A_{\nu}]\, \\
A_{\mu}(\tau,x_i) &\equiv& A_{\mu}^a T^a\, \label{potential} 
\end{eqnarray} 
with $T^a$ denoting the anti-Hermitian generators of $G$ in the fundamental representation. 

The second term in Eq.~(\ref{action}) with $D_{\mu}[\mathcal{A}]=\partial_{\mu}+\mathcal{A}_{\mu}$ represents implementation of the matching condition of Eq.~(\ref{match}) via  $G$-valued Lagrange multiplier field $g^{\mu\nu}$. This field, together with edge modes are considered as an additional set of fields defined on $\partial M$, with outward-pointing unit normal vector $n_{\mu}$ and coordinates $(\tau,x_{\alpha})$, $\alpha=1,2$:  
\begin{eqnarray} 
g_{\mu\nu}(\tau,x_{\alpha}) &\equiv& g_{\mu\nu}^a T^a\,, \qquad (g_{\mu\nu}=-g_{\nu\mu})\,, \\ g(\tau,x_{\alpha}) &=& \exp\!\left(i\phi^a T^a\right)\,. 
\end{eqnarray} 

Furthermore, the boundary term in Eq. (\ref{action}) is required to render the variational principle well defined \cite{Gervais:1976ec}.\footnote{In the infinite-volume clssical theory one sets $\mathcal{A}_{\mu}=0$ \cite{Gervais:1976ec}, which implies that classical gauge fields approach pure-gauge configurations at infinity and thus have finite action. In a finite-volume theory, however, we retain a non-vanishing 'reference' potential, because the relevant on-shell configurations with minimal action and non-trivial topological winding, the instantons, are not pure gauge at the finite boundary. We stress that this technical point has no implication for physical processes mediated by instantons. Indeed, one could alternatively impose pure-gauge boundary conditions within the finite volume at the expense that instanton configurations, while still regular and of finite action, become off-shell \cite{tHooft:1986ooh}.}  We remark, however, that the boundary term in Eq. (\ref{action}) arises from field variations under the implicit assumption that the fields are smooth functions on $M$. If these is not the case, one must introduce further partition of the spacetime manifold to properly treat singular gauge fields \cite{new}.

From the perspective of quantum theory, the partition of the full spacetime $\mathbb{R}^4$ is viewed as a step for infrared regularization of the theory. Indeed, if $M$ is taken to be compact (but not necessarily closed), e.g., a 4D ball of radius $R$, $M\equiv B_R^4$ with a closed boundary $S^3$, the above set-up would define a theory with an infrared cut-off $\propto 1/R$. Upon the relevant regularisation, the boundary must be taken to infinity, $R\to \infty$.    

It is important to emphasise that the boundary fields $\phi^a$ are essential for ensuring invariance of the boundary action under gauge transformations that do not vanish at the boundary (so-called 'large' gauge transformations). For a gauge transformation $h$ with nontrivial boundary value, 
\begin{equation*}
\left.h\right|_{\partial M} \equiv h(\tau,x_{\alpha}) \in G\, 
\end{equation*}
the fields transform as
\begin{eqnarray}
g_{\mu\nu}
&\to&
hg_{\mu\nu}h^{-1}\,,
\nonumber\\
g
&\to&
h g\,,
\nonumber\\
A_{\mu}
&\to&
h A_{\mu} h^{-1}-
h\partial_{\mu} h^{-1}\,.
\label{trans}
\end{eqnarray}
It is straightforward to see that the boundary value of the gauge potential (\ref{match}), as well as the entire action (\ref{action}) is invariant under these transformations. 

The variation of the action (\ref{action}) with respect to $A^a_{\nu}$, $g_{\mu\nu}^a$, and $\phi^a$ yields, in addition to the standard classical bulk equations of motion, the boundary equations of motion (i.e., the boundary conditions):
\begin{align}
 \left. n_{\mu} F^{\mu\nu}\right\vert_{\partial M}&=n_{\mu}g^{\mu\nu} \,, \label{bc1} \\
\partial_{\mu} (g^{\mu\nu} n_{\nu})&= \left[ g^{\mu\nu} n_{\nu}\,,\, g^{-1} D_{\mu}[\mathcal{A}] g \right]\,, \label{bc2} \\
\left.A_{(\tau,\alpha)}\right \vert_{\partial M} &= g D_{(\tau,\alpha)}[\mathcal{A}] g^{-1}\,.
\label{bc3}
\end{align}

A remark on the utility of the edge-mode formalism is in order. At the classical level, the connection between the variational principle and the resulting equations of motion typically assumes one of two conditions: either the fields fall off sufficiently rapidly near the boundary so that boundary terms can be neglected, or the boundary values of the fields are fixed so that their variations vanish on the boundary. In theories with nontrivial topological structure and long-range fields, the first assumption is generally invalid. The second option is also problematic: there may exist no field configuration compatible with the assumed boundary values, or the imposed conditions may select a particular solution while excluding other physically relevant configurations. This issue becomes especially acute in quantum theory, where different field histories contribute to the same physical observable. Introducing boundary conditions dynamically via edge modes resolves these difficulties.

In particular, in the quantisation of topological charge in a finite-volume theory, the bulk topological charge can “leak” through the boundary and is, in general, no longer integer-valued. However, the associated flux induces a change in the state of the edge modes, which restores large gauge invariance at the boundary. As a result, the total topological charge, including contributions from the boundary edge modes, remains an integer. In other words, edge modes on the boundary encode the topological winding beyond the boundary, ensuring that an observer within the bounded spacetime $M$ perceives the full winding of the infinite-volume theory. Therefore, a proper treatment of edge-mode dynamics is essential when analysing quantum tunnelling between states of different topological numbers. This will be demonstrated in the following sections.

\section{Gauge invariance and quantisation of the topological charge}

In the presence of a boundary, the topological charge acquires an additional boundary contribution and takes the form
\begin{equation}
\nu(R)=\int_M\frac{d^4x}{16\pi^2}\,\mathrm{Tr}\left(F_{\mu\nu}\tilde F^{\mu\nu}\right)
-\int_{\partial M}\frac{d^3x}{16\pi^2}\, n_{\mu} K^{\mu}[\mathcal{A}]\,,
\label{topcharge1}
\end{equation}
where
\begin{equation}
K^{\mu}[\mathcal A]=\epsilon^{\mu\nu\rho\sigma}\,\mathrm{Tr}\left(\mathcal{A}_{\nu}\mathcal{F}_{\rho\sigma}
-\frac{2}{3}\mathcal{A}_{\nu}\mathcal{A}_{\rho}\mathcal{A}_{\sigma}\right)
\label{CS}
\end{equation}
is the Chern–Simons current built out of the 'reference' gauge potential $\mathcal{A}_{\mu}$ defined on $\partial M$. Under the boundary condition given in Eq.~(\ref{bc3}) and for smooth bulk field configurations, the topological charge can then be expressed purely as a boundary integral over the edge modes,
\begin{equation}
\nu(R)=\int_{\partial M}\frac{d^3x\,n_{\mu}\epsilon^{\mu\nu\rho\sigma}}{16\pi^2}\mathrm{Tr}\left[
(g^{-1}\partial_{\nu}g)
(g^{-1}\partial_{\rho}g)
(g^{-1}\partial_{\sigma}g)
\right].
\label{topcharge2}
\end{equation}
For edge-mode configurations $g:\,S^3\to G$, this expression is immediately recognised as an integer-valued winding number. Hence, the full information about topology is encoded in the boundary edge modes.

At this point, we stress the importance of maintaining invariance under the large gauge transformations given in Eq.~(\ref{trans}). In particular, the topological index defined in Eq.~(\ref{topcharge2}) is manifestly gauge invariant. If one were to retain only the “standard” bulk term in Eq.~(\ref{topcharge1}), it would be neither gauge invariant nor integer-valued. 

This result is hardly surprising, as topological quantisation is deeply ingrained in non-Abelian gauge theories. The integrality of the topological charge fundamentally follows from the fact that the boundary spacetime is homeomorphic to a closed three-sphere, $\partial M \cong S^3$, regardless of whether the spacetime is infinitely extended or regulated at finite volume. This follows from the observation that any smooth gauge transformation on the boundary\footnote{Note, while 'large' gauge transformations are smooth on the 3D boundary, they cannot be smoothly extended to the 4D bulk.} can be continuously deformed such that it approaches the identity matrix at the boundary $\partial M$. This identification collapses all boundary points of $\partial M$ into a single point, rendering $\partial M$ topologically equivalent to $S^3$. Consequently, field configurations that define maps from $S^3$ to the compact manifold of a simple gauge group $G$ (and the corresponding quantum states) are classified by an integer-valued topological charge associated with the third homotopy group, $\pi_3(G) = \mathbb{Z}$. 

A concrete illustration of the general considerations above is provided by the \(SU(2)\) BPST instanton solution \cite{Belavin:1975fg} in a regular gauge, considered in a finite-volume theory defined on a four-dimensional Euclidean ball \(B^4_R\) of radius \(R\):
\begin{equation}
\bar A_{\mu}(x)=\frac{(x-\bar x)^2}{(x-\bar x)^2+\rho^{2}} u\partial_{\mu} u^{-1}\,,~~ |x-\bar x|<R\,,~ \bar x\in B^4_R,
\label{inst_reg}
\end{equation}
where \(\bar x\) is the position of the instanton within the bounded region, \(\rho\) is its size, and  
\begin{equation}
 u(x)=\frac{(x_{4}-\bar x_4)\,\mathbbm{1}+i\,(x_{a}-\bar x_a)\tau^{a}}{|x-\bar x|}.
\label{wind}
\end{equation}
The bulk contribution to the topological charge (the first term in Eq.~(\ref{topcharge1})) is non-integer and depends on both the instanton location, \(d=|\bar x|\), and its size.\footnote{Both scale and translational invariances are explicitly broken in the presence of a boundary. Hence, the corresponding instanton zero modes are only approximate, i.e. we are dealing with constrained instantons.}
\begin{equation}
\nu_{\rm bulk}(R,d,\rho)\approx
\begin{cases}
\dfrac{R^4\left(R^2+\rho^2\right)}{\left(R^2+\rho^2\right)^3}, & d\ll R, \\[8pt]
\dfrac{1}{2}, & (d-R)\ll\rho\ll R.
\end{cases}
\end{equation}
For large instantons, \(\rho\gg R\), one finds \(\nu_{\rm bulk}\approx 0\). Since \(\nu_{\rm bulk}\) explicitly depends on geometric quantities, its physical interpretation is that of the instanton number density contained within the bounded region, rather than a genuine topological charge.

To ensure that the winding is confined within the finite volume \(r<R\), one may consider the extension of the instanton configuration to the region outside the ball, \(\mathbb{R}^4/ B^4_R\), given by the configuration in the singular gauge:
\begin{equation}
\bar {\mathcal{A}}_{\mu}(x)=-\frac{\rho^2}{(x-\bar x)^2+\rho^{2}} u\partial_{\mu} u^{-1}\,,~~ |x|>R,~\bar x\in B^4_R,
\label{inst_sing}
\end{equation}
with the singular centre placed inside \(B_R^4\). Note that \eqref{inst_sing} is regular outside the region \(B^4_R\) as well as on its boundary \(S^3_R\), that is, in the domain of spacetime where it is defined.

On the boundary \(r=R\), the two gauge fields \eqref{inst_reg} and \eqref{inst_sing} are matched according to Eq.~(\ref{bc3}),
\begin{equation}
\left.\bar A_{\mu}^{a}\right|_{r=R}
=-\left.\bar g^{-1}\,D_{\mu}[\bar{\mathcal{A}}]\,\bar g\right|_{r=R},
\label{gauge_match}
\end{equation}
with the specific edge mode configuration where \(\bar g\) carrying an unit winding:
\begin{equation}
\bar g(x)=\left. u\right\vert_{\partial M}\,.
\label{wind}
\end{equation}
Substituting (\ref{gauge_match}) into Eq.~(\ref{topcharge2}) yields the unit topological charge precisely, provided the instanton is located within the confined region, \(\bar x\in B^4_R\).
 
We emphasise that our construction does not employ fixed pure-gauge boundary conditions, which underlie the standard analysis and have been criticised in Refs.~\cite{Ai:2020ptm, Ai:2024cnp}. Instead, the dynamically determined boundary conditions (\ref{bc1})–(\ref{bc3}) allow flux to cross the boundary. Gauge invariance in the presence of boundaries then necessarily requires the inclusion of boundary edge modes to compensate for this flux, thereby preserving the topological structure of the theory.

The claim of Refs.~\cite{Ai:2020ptm, Ai:2024cnp} that topological charge is quantised only in the infinite-volume limit, where the gauge field becomes pure gauge, is therefore incorrect. It follows from an inconsistent treatment of boundary conditions in finite-volume gauge theory, leading to a gauge-noninvariant and non-integer definition of the topological charge. When boundary edge modes are properly accounted for, topological quantisation holds already at finite volume, and there is no obstruction to summing over topological sectors prior to taking the infinite-volume limit.

\section{Topological transitions and emergence of $\uptheta$}

Now we will show that the reversed order of limits, that is, first taking an infinite-volume limit and then performing an infinite sum over the topological sectors, also leads to a consistent result. In our picture, moving the boundary to infinity does not completely eliminate the boundary edge modes; rather, the dynamics of edge modes are effectively frozen at infinity, but they still carry a topological charge. These charges produce a nontrivial weighting of topological sectors when they are subsequently summed over, leading to the emergence of the $\uptheta$-parameter in physical correlators.

Our goal is to evaluate the transition amplitude (\ref{amp}). In the path integral representation, the integral over the fields must be taken in a given topological sector $\nu$. This can be enforced by inserting $\delta(\nu -\nu(R))\equiv \int_{-\pi}^{\pi}\frac{d\theta}{2\pi}{\rm e}^{i\uptheta \nu(R)}$ in the full functional integral. One thus obtains: 

\begin{eqnarray}
\label{path}
&\langle n-\nu \lvert e^{-H_V T} \rvert n \rangle
= \\
&\int_{-\pi}^{\pi}\frac{d\uptheta}{2\pi}\int DA_{\mu}Dg^{\mu\nu}D\phi^a\,{\rm e}^{-S_{\rm bulk} -S_{\rm bdry}+i\theta \nu(R)+\text{g.f.+gh.}}\,, \nonumber 
\end{eqnarray}
where the classical action is split into bulk and boundary contributions as in Eq. (\ref{action}), and the topological charge $\nu(R)$ is given by Eq. (\ref{topcharge2}). Thus, the $\theta$-term contributes as the Wess-Zumino-Witten term to the boundary action of the edge modes.\footnote{Our treatment closely parallels the description of instantons in terms of a boundary nonlinear $\sigma$-model within the gauge-invariant framework of the vacuum wavefunctional \cite{Brown:1998cp}.
} 

To evaluate the above path integral in the semiclassical approximation, one decomposes the gauge field into a classical instanton configuration and small fluctuations around it. Correspondingly, the edge modes are split into background configuration, which carry an integer topological charge (\ref{topcharge2}), and small fluctuations about these backgrounds. In the Gaussian approximation, the path integrals over bulk and edge fluctuations factorise, except for the integrations over the instanton translational and dilatational zero modes. As a result, a complete evaluation of the amplitude (\ref{path}) in finite volume remains technically involved, even in the non-interacting instanton gas approximation. However, the key point to observe is that the effective dynamics of 3D boundary edge modes entirely determine the topological features of the transition.  

If one instead takes the infinite 4-volume limit first, \(R \to \infty\), the effective kinetic term for the edge modes, $\propto R\int d^3x {\rm Tr}\left(g\partial_{\mu} g^{-1}\right)\left(g\partial^{\mu} g^{-1}\right)$, would diverge, rendering the modes non-dynamical in this limit. What survives is solely their contribution to the topological \(\uptheta\)-term through the background edge-mode configuration, which, as discussed above, ensures that \(\nu(R)\) is integer-valued. At the same time, the bulk instanton action approaches its standard value \(8\pi^{2}|\nu|/g^{2}\), since the bulk contribution to the topological charge also becomes an integer, $\lim_{R\to\infty}\nu_{\rm bulk}\in \nu \in \mathbf{Z}\,$. In this way, one recovers the standard \(\nu\)-dependence of the amplitude (\ref{amp}) through the \(\uptheta\)-term in (\ref{path}).

\section{The fate of $\uptheta$-vacua in the presence of fermion}. 

Fermions in an instanton background exhibit spectral asymmetry through Euclidean zero-energy modes that carry an unbalanced anomalous global charge. The charge associated with these vacuum modes is precisely determined by the instanton’s topological charge, $\nu$, in accordance with the index theorem \cite{Atiyah:1963zz}.

This has profound consequences for the $\uptheta$-vacuum structure of the theory. The anomalous global charge carried by the fermion zero modes manifests itself as a vacuum condensate of a composite fermion operator, the ’t Hooft vertex. The collective phase of this condensate then emerges as a propagating degree of freedom, dynamically linking previously disconnected $\uptheta$-sectors.

The spontaneous breaking of an anomalous global symmetry is realised not in the conventional Goldstone mode, but rather in a Higgs regime \cite{Dvali:2005an, Dvali:2005ws}. This occurs because the anomaly induces a coupling between the emergent would-be Goldstone boson and a long-range composite three-form gauge field, generating a $BF$-type topological mass term.

This phenomenon is very generic and fundamentally rooted in the topological structure of gauge theory. It arises in confined theories such as QCD (through the $\eta'$ meson), in theories in the Higgs phase, most notably in the electroweak sector of the Standard Model \cite{Dvali:2024zpc, Dvali:2025pcx}, and (super)gravity \cite{Dvali:2024dlb}. 
  
\section{Conclusion}
In this paper, we scrutinise the claim made in Refs.~\cite{Ai:2020ptm, Ai:2024cnp} that theories with a $\uptheta$-vacuum structure, most notably quantum chromodynamics, exhibit no physical 
$\uptheta$-term and hence no CP violation. We have shown, contrary to these assertions, that an integer-valued topological charge can be rigorously defined already in a finite-volume theory. This is achieved by formulating the theory with open boundary conditions supplemented by dynamical boundary degrees of freedom, the edge modes. The crucial role of these edge modes is to restore invariance under large gauge transformations, which would otherwise be violated in the presence of a boundary. At the same time, they ensure the integrality of the properly defined topological charge. In this framework, topological charge and its fluctuations are correctly attributed to the dynamics of the boundary edge modes.

Having established a consistent finite-volume formulation, we then addressed the issue of the order of limits, which constitutes a central criticism of the standard calculations in Refs.~\cite{Ai:2020ptm, Ai:2024cnp}. In our construction, taking the infinite-volume limit renders the edge modes non-dynamical. Nevertheless, the topological information carried by their background configurations survives this limit and precisely reproduces an integer topological charge, entering physical correlators through the CP-violating $\uptheta$-term.

In summary, we conclude that the claim of $\uptheta$-independence of physical correlators, and the associated absence of CP violation, is incorrect. The underlying error in Refs.~\cite{Ai:2020ptm, Ai:2024cnp} stems from an improper treatment of boundary conditions in finite-volume gauge theory and a consequent misinterpretation of the infinite-volume limit. 

In the presence of fermions carrying an anomalous charge, a scalar degree of freedom generically emerges that links previously disconnected $\uptheta$-sectors via the spontaneous breaking of the anomalous symmetry. If this symmetry is exact at the classical level, the $\uptheta$-parameter ceases to be observable.

In the context of QCD, this scenario is realised when the up-quark is massless, in which case the $\eta'$ meson dynamically removes the $\uptheta$-term. \cite{Dvali:2005an}. In the electroweak theory, the relevant anomalous symmetry is $(B+L)$. The electroweak $\uptheta$-term is removed by a yet-to-be-discovered emergent particle state, referred to as the electroweak $\eta_{\rm w}$.

\section*{Acknowledgments}
 Some details of this work were shaped in discussions with Lasha Berezhiani, Gia Dvali and Oto Sakhelashvili during my sabbatical stay at the Max Planck Institute for Physics in Munich. I am indebted to their input and hospitality. I would also like to thank Carlos Tamarit for communicating his ideas during the University of Melbourne TPP online seminar, and Prateek Agrawal, Latham Boyle, Zurab Berezhiani, Fawad Hassan, Roberto Contino, Elden Loomes and Alessandro Strumia for useful discussions on various aspects of $\uptheta$-vacua and the strong CP problem. The work was partially supported by the Australian Research Council Discovery Projects grant DP220101721.



\end{document}